\documentstyle[aps,epsf]{revtex}
\begin{document}
\title{Fractal Dimension of Julia Set for Non-analytic Maps}
\author{Chao Tang}
\address{NEC Research Institute, 4 Independence Way, Princeton, New
Jersey 08540}

\maketitle

\begin{abstract}
The Hausdorff dimensions of the Julia sets for non-analytic maps: $f(z) =
z^2 + \epsilon z^*$ and $f(z) = {z^*}^2 + \epsilon$ are
calculated perturbatively for small $\epsilon$. It is shown that Ruelle's
formula for Hausdorff dimensions of analytic maps can not be generalized
to non-analytic maps.
\end{abstract}

\section{Introduction}
The Julia set $J$ of a map is the closure of the unstable periodic points
\cite{jul,bro,fat,dev}. It is an invariant set of the map and is usually a
``repeller'', that is, points close to $J$ will be repelled away by
successive iterations of the map. A simple example is the map on the
complex plane: $f(z)=z^2$, for which $J$ is the unit circle. Points close
to $J$ will flow to one of the two stable fixed points: 0 and $\infty$.
Thus $J$ is the boundary or separator of basins of attraction. A much more
complicated geometry appears for the Julia set of the map: 
\begin{equation}
f(z)=z^2 + c, 
\label{z2c}
\end{equation}
where $c$ is a non-zero constant (see Fig.~1(a) for an example and 
Ref.~\cite{dev} for many other examples). In this case, $J$ is a fractal
and its topology undergoes drastic changes as $c$ varies.
\begin{figure}
\vskip -2.0true cm
\centerline{\epsfxsize=5in
\epsffile{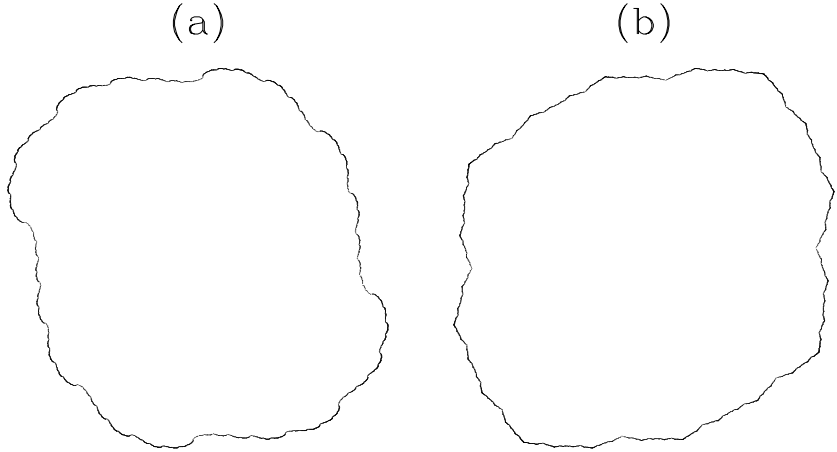}}
\vskip -3.0true cm
\caption{The Julia set for (a) $f(z)=z^2+\epsilon$ and (b) $f(z)={z^*}^2
+\epsilon$, for $\epsilon= 0.15 + i 0.15$.}
\end{figure}

Before we proceed further, let us define a few notations. We denote
$f^n$ to be $n$ successive iterations of the map. That is
$f^n(z)=f(f^{n-1}(z))$. The
set of all unstable cycles of length $n$ is denoted by 
${\rm Fix} \, f^n$. $Df$ is the derivative matrix of $f$. If $f$ is an
analytic map, i.e. $\partial u/\partial x=\partial
v/\partial y$ and $\partial v/\partial x=-\partial u/\partial y$ with
$f(z=x+iy)=u+iv$, then $\det Df=|df/dz|^2$.

For analytic maps, the Hausdorff dimension $D_H$ of the Julia set $J$
can be calculated with a formula due to a theorem of Ruelle \cite{rue}:
\begin{equation}
\lim_{n \to \infty} A_n(D_H) = 1,
\label{an1}
\end{equation}
where
\begin{equation}
A_n(D) = \sum_{z \in {\rm Fix} f^n} |\frac{df^n}{dz}|^{-D}.
\label{an}
\end{equation}
Using the formula, Ruelle \cite{rue} and Widom {\it et al.} \cite{mike}
calculated $D_H$ for the map (\ref{z2c}) in powers of $c$ for small
$|c|$. It was not clear then whether the formulas (\ref{an1}) and (\ref{an})
can be generalized to non-analytic maps. The natural generalization of
(\ref{an}) to non-analytic maps would be
\begin{equation}
A_n(D) =  \sum_{z \in {\rm Fix} f^n} |\det Df^n|^{-D/2}.
\label{an-na}
\end{equation}
The calculations I present below show that the combination of 
(\ref{an1}) and (\ref{an-na}) does not give the correct $D_H$ for 
non-analytic maps in general and $D_H$ can be calculated directly
with the perturbation theory developed in Ref.~\cite{mike}.

\section{The map \lowercase{$f(z)=z^2 + \epsilon z^*$}}
Let us first consider the non-analytic map
\begin{equation}
f(z)=z^2 + \epsilon z^*,
\label{na1}
\end{equation}
where $*$ denotes the complex conjugate.
When $\epsilon=0$ the Julia set $J$ is the unit circle and can be
parametrized as $z(t)=e^{2\pi it}$. The map on $J$ is
\begin{equation}
f(z(t)) = z(2t).
\label{fzt}
\end{equation}
When $\epsilon \neq 0$ but small enough so that $J$ is topologically
equivalent to a circle we can still parametrize $J$ so that
Eq.~(\ref{fzt}) is satisfied \cite{bro,fat}.
If a map $f_\epsilon$ with a parameter $\epsilon$ satisfies
\begin{equation}
[f_\epsilon(z)]^* = f_{\epsilon^*}(z^*),
\label{commu}
\end{equation}
then
\begin{equation}
z \in {\rm Fix} \, f_\epsilon^n \ \ \Leftrightarrow \ \ z^* \in {\rm Fix} \,
f_{\epsilon^*}^n,
\end{equation}
which implies that
\begin{equation}
J(f_\epsilon) = [J(f_{\epsilon^*})]^*,
\label{jj}
\end{equation}
where $J(f)$ is the Julia set of $f$. In particular, if $J$ can be
parametrized as $z(t)$ then 
\begin{equation}
z_\epsilon(t)=z_{\epsilon^*}^*(-t).
\label{zz}
\end{equation}
It is easy to see that the map (\ref{na1}) satisfies Eq.~(\ref{commu}).

Following Widom {\it et al.}
\cite{mike}, we formally expand $z(t)$ in powers of $\epsilon$
\begin{equation}
z(t)=e^{2\pi it}[1+\epsilon U_1(t) + \epsilon^*\tilde{U_1}(t) +
\epsilon^2 U_2(t) + {\epsilon^*}^2 \tilde{U_2}(t) +
\epsilon\epsilon^* \hat{U_2}(t) + \cdots],
\label{expand}
\end{equation}
where the functions $U_1(t), \tilde{U_1}(t), U_2(t), \tilde{U_2}(t),
\hat{U_2}(t), \dots$ are all periodic with period 1. Eq.~(\ref{zz})
implies that all the functions $U(t)$ satisfies $U(t)=U^*(-t)$. Substituting
(\ref{expand}) into (\ref{fzt}) and equating terms with the same power
of $\epsilon$, we get
\begin{eqnarray}
U_1(2t)-2U_1(t) & = & e^{-6\pi it}, \\
\tilde{U_1}(2t)-2\tilde{U_1}(t) & = & 0, \\
U_2(2t)-2U_2(t) & = & U_1^2(t)+e^{-6\pi it} \tilde{U_1}^*(t), \\
\tilde{U_2}(2t)-2\tilde{U_2}(t) & = & \tilde{U_1}^2(t), \\
\hat{U_2}(2t)-2\hat{U_2}(t) & = & 2U_1(t)\tilde{U_1}(t)+e^{-6\pi it}U_1^*(t).
\label{eq_u}
\end{eqnarray}
The solutions are
\begin{eqnarray}
U_1(t) & = & - \sum_{k=1}^{\infty} \frac{e^{-3\pi i2^kt}}{2^k}, 
\label{u1} \\
\tilde{U_1}(t) & = & 0, \label{u2} \\
U_2(t) & = & -\sum_{j,k,l=1}^{\infty} \frac{e^{-3\pi
i2^j(2^{k-1}+2^{l-1})t}} {2^{j+k+l}}, \label{u3} \\
\tilde{U_2}(t) & = & 0, \label{u4} \\
\hat{U_2}(t) & = & \sum_{j,k=1}^{\infty} \frac{e^{3\pi i
2^j(2^{k-1}-1)t}}{2^{j+k}}. 
\label{u5}
\end{eqnarray}
It is easy to see from Eq.~(\ref{fzt}) that unstable cycles of length
$n$ are 
\begin{equation}
{\rm Fix} \, f^n = \{ z(t_j): t_j=\frac{j}{2^n-1}, j=0,1,\dots,2^n-2 \}. 
\label{fixfn}
\end{equation}

We now evaluate $A_n(D)$ as defined in (\ref{an-na}). Note that
\begin{eqnarray}
\det Df^n & = & \prod_{i=0}^{n-1} \det 
      \left( \begin{array}{cc}
  2x_i+{\rm Re}(\epsilon)\  &  -2y_i+{\rm Im}(\epsilon)  \\
  2y_i+{\rm Im}(\epsilon)\  &   2x_i-{\rm Re}(\epsilon)
      \end{array} \right) \nonumber \\
& = & \prod_{i=0}^{n-1} (4z_i^2 - |\epsilon|^2) \nonumber \\
& = & 4^n (1-\frac{|\epsilon|^2}{4})^n \prod_{m=0}^{n-1} |z(2^mt_j)|^2,
\nonumber
\end{eqnarray}
where the last equality holds to the second order in $\epsilon$.
Denote
\begin{equation}
<G(t)>_n = \frac{1}{2^n-1} \sum_{j=0}^{2^n-2} G(t_j),
\label{gt}
\end{equation}
where $t_j$'s are given by Eq.~(\ref{fixfn}).
\begin{eqnarray}
A_n(D) & = & \sum_{z\in{\rm Fix} f^n}|\det Df^n|^{-D/2} \nonumber \\
       & = & 2^{-Dn}(2^n-1)(1-\frac{|\epsilon|^2}{4})^{-Dn/2}
<\prod_{m=0}^{n-1} |z(2^mt_j)|^{-D}>_n.
\label{an-mid}
\end{eqnarray}
Substituting Eqs.~(\ref{u1})-(\ref{u5}) into (\ref{expand}) and using the
identity
\begin{equation}
<e^{2\pi imt}>_n = \left\{ \begin{array}{ll}
             1,   \quad         & m = 0 \bmod {2^n-1} \\
             0,   \quad         & m \neq 0 \bmod {2^n-1}
                           \end{array}
                   \right.
\label{id}
\end{equation}
it can be shown, after some algebra, that
\begin{equation}
<\prod_{m=0}^{n-1} |z(2^mt_j)|^{-D}>_n = 1+|\epsilon|^2(\frac{D^2n}{4} -
\frac{Dn}{2} - \frac{D^2}{2} - \frac{Dn}{2^{n+1}} +
\frac{D^2n}{2^{n+3}}), \quad (n>2).
\label{av-z}
\end{equation}
Substituting (\ref{av-z}) into (\ref{an-mid}) yields
\begin{equation}
A_n(D) = 2^{n(1-D)} [1+|\epsilon|^2(\frac{D^2n}{4}-\frac{3Dn}{8})],
\quad (n>>1).
\label{and}
\end{equation}
If we were to use Eqs.~(\ref{and}) and (\ref{an1}) to obtain a Hausdorff
dimension, we would get $D_H=1-|\epsilon|^2/(8\ln2)$, a value smaller than
1 for small but nonzero $\epsilon$. We show in the following that this
value of $D_H$ is incorrect.

Let
\begin{equation}
\chi_n(D) = \sum_{j=0}^{2^n-2} \frac{|z(t_{j+1})-z(t_j)|^D}{(2\pi)^D},
\label{xnd}
\end{equation}
where $z(t_j) \in {\rm Fix} \, f^n$ (Eq.~(\ref{fixfn})). The Hausdorff
dimension $D_H$ of the set ${\rm Fix} \, f^n$ in the limit $n \to \infty$ is
such that
\begin{equation}
\lim_{n\to\infty} \chi_n(D_H) =1.
\label{xd1}
\end{equation}
This $D_H$ should also be the $D_H$ of the Julia set $J$. 
We now evaluate $\chi_n(D)$ to the second order in $\epsilon$. Putting
Eqs.~(\ref{u1})-(\ref{u5}) into Eq.~(\ref{expand}), we write
\begin{equation}
z(t_{j+1})-z(t_j) = C_0 + C_1 |\epsilon| + C_2 |\epsilon|^2.
\label{dz}
\end{equation}
Then to the second order in $\epsilon$,
\begin{eqnarray}
\chi_n(D) =&\frac{|C_0|^D}{(2\pi)^D}& (2^n-1)\{1 + \frac{D|\epsilon|}
{|C_0|^2} <{\rm Re}(C_0^*C_1)>_n  \nonumber \\
&+&\frac{D|\epsilon|^2}{|C_0|^2}\,
[\frac{1}{2}<|C_1|^2>_n + <{\rm Re}(C_0^*C_2)>_n +
\frac{D-2}{2|C_0|^2} <({\rm Re}(C_0^*C_1))^2>_n] \},
\label{chi-long}
\end{eqnarray}
where Eq.~(\ref{gt}) is used. With the help of the identity
(\ref{id}) we get
\begin{eqnarray}
|C_0|^2 & = & 2(1-\cos\frac{2\pi}{2^n-1}), \label{c0} \\
<{\rm Re}(C_0^*C_1)>_n & = & 0, \quad (n>2) \\
<|C_1|^2>_n & = & F(n), \\
<{\rm Re}(C_0^*C_2)>_n & = & \frac{|C_0|^2}{2} (1+\frac{1}{2^n}), \\
<({\rm Re}(C_0^*C_1))^2>_n & = & \frac{1}{2} |C_0|^2 <|C_1|^2>_n,
\label{c0c1}
\end{eqnarray}
where
\begin{equation}
F(n) = \frac{2}{3} - 2 \sum_{k=1}^{\infty}\frac{1}{4^k} \cos 2\pi
\frac{3\cdot2^{k-1}-1}{2^n-1}.
\end{equation}
The function $F(n)$ can easily be solved for large $n$ in the following
way. Note that for $n>>1$
\begin{eqnarray}
F(n+1) & = & \frac{1}{2} (1-\cos\frac{3\cdot2\pi}{2^{n+1}}) +
\frac{1}{4} F(n) \nonumber \\
& = & \frac{9\pi^2}{4^{n+1}} + \frac{1}{4} F(n).
\label{fn}
\end{eqnarray}
Substituting $F(n)=H(n)/4^n$ into Eq.~(\ref{fn}), we have
\begin{equation}
H(n+1) = 9\pi^2 + H(n),
\end{equation}
which has the solution
\begin{equation}
H(n) = 9\pi^2 n + a,
\label{hn}
\end{equation}
where $a$ is a constant independent of $n$. From Eqs.~(\ref{chi-long}),
(\ref{c0}) - (\ref{c0c1}), and (\ref{hn}),
\begin{eqnarray}
\chi_n(D)&=&2^{n(1-D)}[1+ |\epsilon|^2 (\frac{D}{2}+
\frac{D}{4}\frac{<|C_1|^2>_n}{|C_0|^2})] \nonumber \\
&=&2^{n(1-D)}(1+ \frac{9}{16}nD^2|\epsilon|^2), \quad (n>>1).
\label{chin}
\end{eqnarray}
Eqs.~(\ref{xd1}) and (\ref{chin}) imply 
\begin{equation}
D = 1 + \frac{9}{16\ln 2} |\epsilon|^2.
\label{dh1}
\end{equation}

\section{The Map \lowercase{$f(z)={z^*}^2+\epsilon$}}
Next, we consider the non-analytic map
\begin{equation}
f_\epsilon (z)={z^*}^2+\epsilon.
\label{na2}
\end{equation}
The map (\ref{na2}) has the property of Eq.~(\ref{commu}), so that
$J(f_\epsilon) = [J(f_{\epsilon^*})]^*$.

Let us parametrize $J$ in such a way so that
\begin{equation}
f(z(t)) = z(-2t), \quad z(t) \in J.
\end{equation}
The set of unstable cycles of length $n$ is
\begin{equation}
{\rm Fix} \, f^n = \{ z(t_j): t_j=\frac{j}{(-2)^n-1}, j=0,\pm 1,\pm 
2,\dots \}.
\end{equation}
The number of elements in ${\rm Fix} \, f^n$ is $|(-2)^n-1|$.
Following similar procedures as in the previous section, we have
\begin{eqnarray}
U_1(t) & = & - \sum_{k=1}^{\infty} \frac{e^{-2\pi i4^kt}}{4^k}, \\
\tilde{U_1}(t) & = & -2\sum_{k=1}^{\infty} \frac{e^{-\pi i4^kt}}{4^k}, \\
U_2(t) & = & -6\sum_{j,k,l=1}^{\infty} \frac{e^{-2\pi
i4^j(4^{k-1}+4^{l-1})t}} {4^{j+k+l}}, \\
\tilde{U_2}(t) & = & -12\sum_{j,k,l=1}^{\infty} \frac{e^{-\pi
i4^j(4^{k-1}+4^{l-1})t}} {4^{j+k+l}} + \sum_{k,l=1}^{\infty}
\frac{e^{-\pi i(4^k+4^l)t}} {4^{k+l}}, \\
\hat{U_2}(t) & = & -4\sum_{j,k,l=1}^{\infty} \frac{e^{-\pi i
2^j(4^k+4^l/2)t}}{2^{j+2k+2l}}.
\end{eqnarray}
$A_n(D)$ (Eq.~(\ref{an-na})) and $\chi_n(D)$ ((Eq.(\ref{xnd})) can be
calculated to be
\begin{equation}
A_n(D)=\chi_n(D) = 2^{n(1-D)} (1 + \frac{1}{4} n D^2 |\epsilon|^2).
\label{xn-na2}
\end{equation}
In this case, $A_n(D)=\chi_n(D)$ and it gives the correct Hausdorff
dimension
\begin{equation}
D_H = 1 + \frac{|\epsilon|^2}{4\ln 2}.
\label{dh-na2}
\end{equation}
The reason for Ruelle's formula to work in this case is that for the
non-analytic map (\ref{na2}) $f^2(z)$ is analytic:
\begin{equation}
f^2(z)=(z^2+\epsilon^*)^2+\epsilon,
\end{equation}
and that $J(f^2)=J(f)$.  Note that (\ref{dh-na2}) is the same as the
$D_H$ of the analytic map (\ref{z2c}) $f(z)=z^2+\epsilon$ \cite{rue,mike},
to the second order in $\epsilon$.  Indeed, $f^2(z)$ and thus $J$ are
identical for the two maps (\ref{z2c}) and (\ref{na2}) for real $\epsilon$.
For complex $\epsilon$, however, the two Julia sets look quite different
(Fig.~1).

\section{Discussion}
Since Ruelle's formula (\ref{an1}) relies on the analyticity of the map,
it is no surprise that it brakes down for non-analytic maps.  When $J$ is
a closed curve, $D_H$ can be calculated from $\chi_n(D)$ (Eq.~(\ref{xnd}))
for both analytic and non-analytic maps.  When $J$ is no longer
topologically a circle, it can be difficult to utilize a formula based on
distances between unstable cycle elements.  In this case, it remains a
challenge to formulate an efficient method for the calculation of $D_H$
for non-analytic maps. Finally, the quantity $A_n(D)$ (Eq.~(\ref{an-na}))
can be very useful even for non-analytic maps. For example, it can be
used to calculate the escape rate for points close to $J$ \cite{kt}.

\end{document}